\documentclass[12pt, a4paper]{article}
\usepackage{lineno}
\pdfoutput=1
\usepackage{color}
\usepackage{graphicx}
\usepackage{caption}
\usepackage{amsmath}
\usepackage{subcaption}
\usepackage{mathtools}
\usepackage{url}
\usepackage[font=scriptsize,labelfont=bf,skip=6pt]{caption}
\usepackage[T1]{fontenc}
\usepackage{authblk}

\begin{document}
\title{Performance of the triple GEM detector built using commercially manufactured GEM foils in India}
\author{Mohit Gola}
\author{Aashaq Shah}
\author{Asar Ahmed}
\author{Shivali Malhotra\thanks{Corresponding Author: shivali.malhotra@cern.ch}}
\author{Ashok Kumar}
\author{Md. Naimuddin}
\affil{Department of Physics $\&$ Astrophysics, University of Delhi, Delhi, India}
\maketitle
\begin{abstract}
The Gas Electron Multiplier (GEM) detectors has been utilized for various applications due to their excellent spatial resolution, high rate capabilities and flexibility in design. The GEM detectors stand as a promising device to be used in nuclear and particle physics experiments. Many future experiments and upgrades are looking forward to use this technology leading to high demand of GEM foils. Until now, CERN is the only reliable manufacturer and distributor of GEM foils, but with technology transfer, few other industries across the globe have started manufacturing these foils employing the same photo-lithographic technique. The Micropack Pvt. Ltd. is one such industry in India which produced first few $10~cm ~\times~ 10~cm$ GEM foils, which were then distributed to few collaborating partners for testing reliability and performance of foils before they can be accepted by the scientific community. Characterization of three such foils have already been performed by studying their optical and electrical properties. Using these foils a triple GEM detector has been built and various performance characteristics have been measured. In this paper, we specifically report measurements on gain, resolution and response uniformity, by utilizing local quality control set-ups existing at University of Delhi.
\end{abstract}


\section{Introduction}
The Gas Electron Multiplier (GEM) was first introduced by Fabio Sauli in 1997 \cite{Sauli}, which is a composite grid consisting of two conducting layers separated by a thin insulator (i.e. Kapton/Apical) etched with a regular matrix of open channels \cite{Sauli2}. Detectors built using these foils prove to be one of the most promising particle detectors in various scientific fields such as nuclear and particle physics, astronomy as well as medical diagnostics \cite{TERA, PCT}. This is due to their excellent position resolution \cite{Spatial}, good timing \cite{Timing}, high rate detection capabilities \cite{Rate}, low ion backflow \cite{IBF}, design flexibility and large area coverage \cite{LargeArea}. An increase in beam energy and luminosity in various accelerator facilities generated a lot of interest to use GEM detectors for future experiments or upgrades of existing experiments \cite{TDR}. This increased demand for GEM foils was difficult to be fulfilled solely by CERN, which led CERN to transfer its technology of manufacturing GEM foils to commercial partners. 
An India-based company, Micropack in Bengaluru have started manufacturing GEM foils using CERN's patented double-mask (or bi-conical) etching processes \cite{DblMask}. A detailed description of the fabrication process can be found in \cite{Hoch}. Typically, the insulator thickness is around 50 $\mu m$ which is covered with 5 $\mu m$ Copper layer on both the sides. To start with, $10~cm \times 10~cm$ GEM foils having $\sim$600,000 holes with an inner hole diameter of 50 $\mu m$, outer hole diameter of 70 $\mu m$ and pitch (i.e. the distance between the center of two neighbouring holes) of 140 $\mu m$ have been manufactured (shown in fig.~\ref{fig:Foil}(a)). These foils have been distributed for testing at few institutes within India, including University of Delhi (DU). We have performed the geometrical studies along with the leakage current measurement of these foils for short and long durations \cite{Foils}. In this paper, we report on the performance of a GEM detector assembled using these first commercially manufactured foils in India.

\section{Description of triple GEM detector}

\begin{figure}[!ht]
    \begin{subfigure}[b]{0.46\textwidth}
        \includegraphics[scale=0.2]{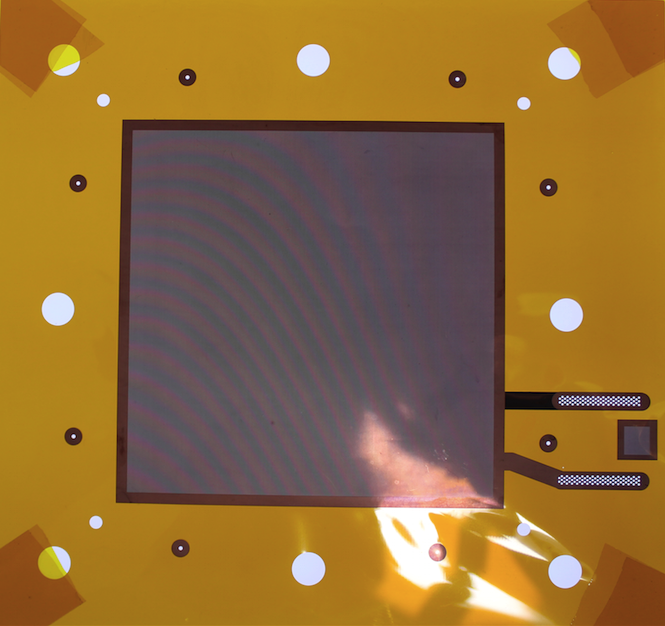}\qquad
        \caption{ }
    \end{subfigure}
    \hspace{-1cm}
    \begin{subfigure}[b]{0.46\textwidth}
        \includegraphics[scale=0.2]{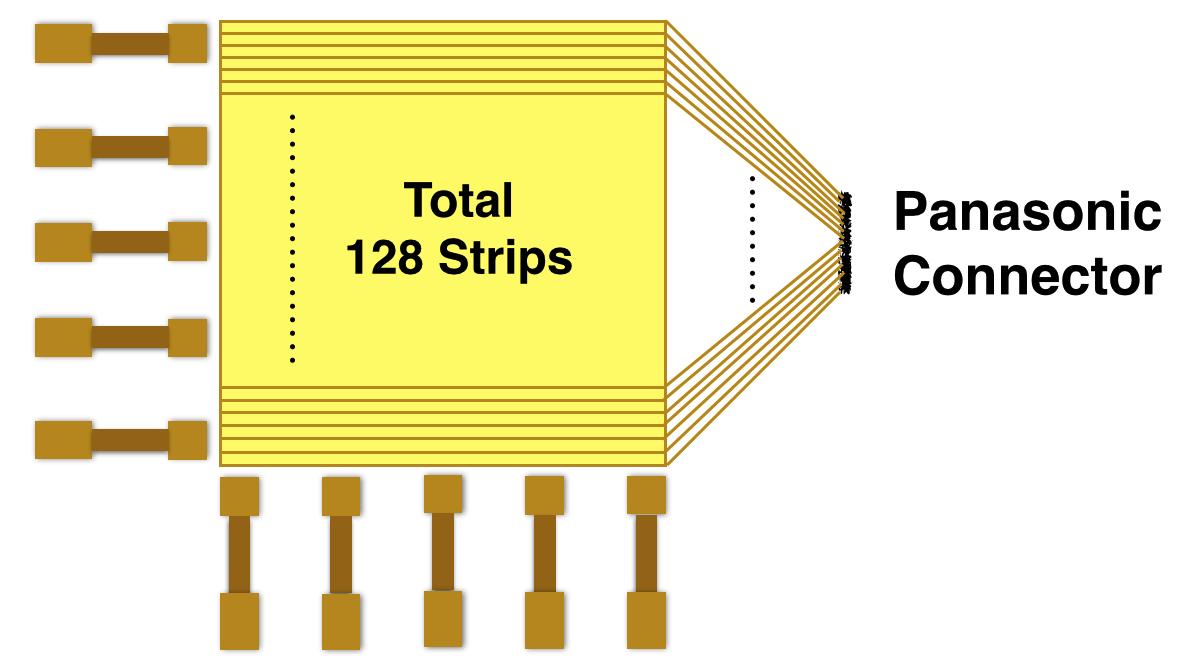}
        \caption{ }
    \end{subfigure}
   \caption{(a) One of the GEM foil manufactured by Micropack Pvt. Ltd. with two HV pads at bottom right, and (b) Schematic of readout board used having 128 strips connected to a Panasonic to LEMO Connector} 
\label{fig:Foil}
\end{figure}

Normally, a triple GEM gaseous detector is built using a drift cathode, three GEM foils, and Printed Circuit Board (PCB) anode (or readout board as shown in fig.~\ref{fig:Foil}(b)), which are mounted in a cascade inside a closed epoxy frame \cite{Constr}. This frame is used for gas tightness having openings located at two diagonal corners, one for gas inlet and another one for gas outlet. In general, there are no exclusions on the choice of gas mixtures \cite{GasProp} for operation of any gaseous detector but to detect ionizing particles and radiations noble gases are best candidates due to their shell configuration. Usually, GEM detector works in a gas medium in which gas ionization takes place due to the electron acceleration under the influence of electric field. Since, working under purely noble gas medium can cause the avalanche creation beyond the limits which leads to sparks and cause a permanent damage to the detector, we also require lower proportion of a quenching gas in the medium. Due to the electro-negativity of Carbon-dioxide (CO$_{2}$), electrons are attracted and attached to its molecules, therefore it works as a quencher in the gas mixture. Many studies have been performed on the selection of gas mixture to be used for this detector (details can be found in ref. \cite{Gases}). We have used the gas mixture with Argon (70$\%$) and CO$_{2}$ (30$\%$)  for our studies \cite{GasOperation} presented throughout in this paper.

\begin{figure}
\centering
\includegraphics[scale=0.4]{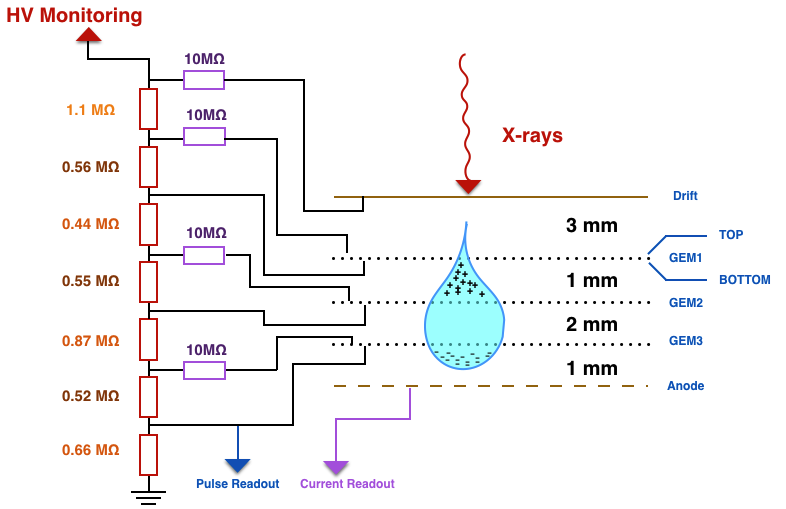}
\caption{Triple GEM detector prototype with high voltage resistive divider, showing various gap configuration and avalanche formation due to incident X-rays.}
\label{fig:Avalanche}
\end{figure}

A standard triple GEM detector prototype \cite{GEMPhysics} has a drift gap, two transfer gaps and an induction gap of $3mm$, $1mm$, $2mm$, and $1mm$ respectively as shown in  fig.~\ref{fig:Avalanche}. This is the same gap configuration which is used in CMS GEM community for muon-endcap upgrade (also called GE1/1) \cite{TDR}. These gaps are created by placing spacers at the edges of the adjacent GEM foils. The drift region is where the primary electrons are generated and due to the field present in this space, the primary charges move towards first GEM foil. This drift field depends on the potential between the cathode and top metal of first GEM foil. The electric field generated inside the holes is $\sim$100 times the field present in the drift region, where each hole work as an individual proportional counter. Due to the field inside the holes, electrons gain rapid energy which leads to an avalanche creation. The electrons further created are accelerated towards the second GEM foil due to the transfer field present between the bottom of one GEM foil and the top of next one. This process is repeated for the succeeding foil with asymmetric potential across the foils and gaps. At last, the induction field, which is present between the last GEM foil and the ground anode, is responsible to deposit all the charges created in the avalanche on the anode plane \cite{ChargeMech}.

Each GEM foil is powered using voltage divider network through a single high voltage (HV) channel (as shown in fig.~\ref{fig:Avalanche}). Since each GEM foil is working at nominal gain, the higher effective gain can be reached at lower working voltages. It also lowers the chances of discharge~\cite{Disc} inside the active medium. The readout strips detect the charge deposition of the amplified signal and transfer it to the data acquisition system for further signal processing.

\section{Experimental setup}
\begin{figure}
\centering
\includegraphics[scale=0.3]{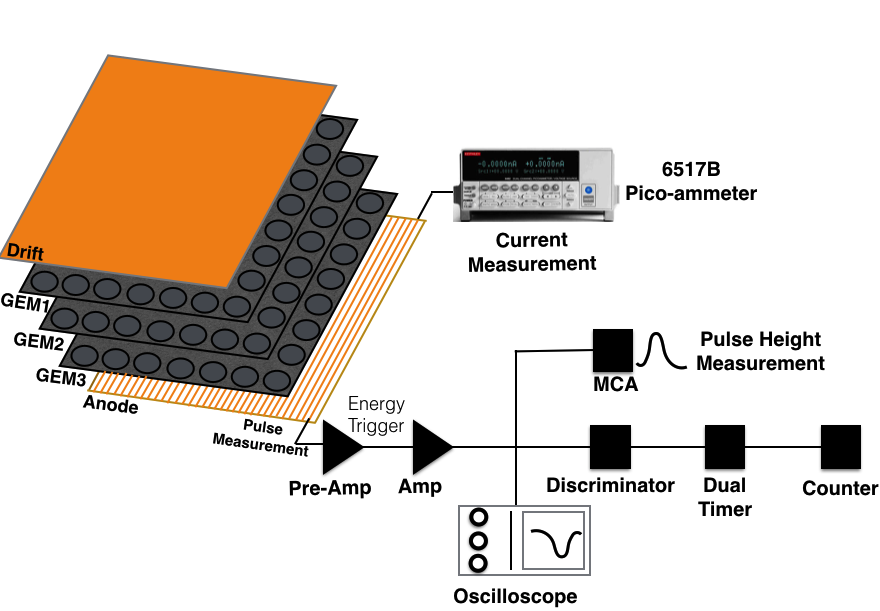}
\caption{Schematic of the set-up used for various studies using $10~cm~ \times ~10~cm$ triple GEM detector prototype.}
\label{fig:Setup}
\end{figure}

In order to study the characteristics of GEM detector, a set-up as shown in fig.~\ref{fig:Setup} have been put in place. The standard gas mixture with 70$\%$ Argon and 30$\%$ CO$_{2}$ is used and flow rate of the gas is kept at 3.0 L/h which is controlled by gas mixing unit with calibrated mass-flow controllers. A readout plane (anode) consisting of 128 strips is used, which is large enough to cover the whole charge cluster. A X-ray source with silver target having characteristic energy of 22.1 $keV$ is used as a generator for the studies mentioned in this paper.

The current measurement is performed using the Panasonic to LEMO connector from the readout and fed into Electrometer 6517B \cite{PicoAmp} which is further interfaced using LabView software.

For the pulse measurement, the signal generated at the anode is processed using charge sensitive pre-amplifier (ORTEC-142IH) and further connected with the Timing Filter Amplifier (TFA-ORTEC 474) to amplify and record the signals obtained from the readout strips. The output of the TFA is fed into the discriminator (CFD-ORTEC 935) and rectangular discriminated pulse is sent to the scalar $\&$ counter for the rate measurement study. 

For measurement of energy resolution, multi-channel analyzer (MCA-ORTEC 927) is used. Since MCA requires positive pulse hence the polarity of the signal is inverted using TFA. This set-up has already been utilized for performing various studies on different prototypes of GEM Detector at DU \cite{DAE}.

\section{Detector characteristics}
After stabilization of the gas flow and reducing the electronic noise, the performance of the triple GEM detector has been studied. The detector is powered using the HV divider that provides an appropriate voltage for safe and stable operation. Additionally, few 10 M$\Omega$ resistors (known as protection resistors) are used between the HV divider and top of each GEM foil to prevent the excess current flow through the foil in case of discharge. The behaviour of the detector is studied by varying the high voltage which results in variation of an electric field across the GEM foils as well as between the four gaps within the detector. The first test, as explained in Section 4.1, was performed to check the detector response to the high voltage. The gain calibration test is explained in Section 4.2, which tells us about the gain at each input voltage and verify whether there are any non-linearities in the performance of the detector. A way to measure the energy resolution is described in Section 4.3, using pulse height spectra obtained from X-ray source. Section 4.4 explains the method used for getting information on uniformity of detector. We have also performed correlation study between the gain uniformity and various defects present in the foil \cite{Foils} in Section 4.5.

\subsection{Behaviour under high voltage}
In order to avoid any damage to readout electronics as well as detector, it is very important to ascertain the behaviour of the detector under high voltages. Before starting the measurement, a total resistance of the circuit was measured to be 5 M$\Omega$, which includes 4.7 M$\Omega$ from HV divider and 0.3 M$\Omega$ from HV filter which is placed between the power supply and detector. According to the standard procedure used by CMS GEM community \cite{DAEProc}, GEM detector can be operated upto 4900 Volts while flushing non-amplifying gas such as CO$_{2}$. Therefore, the detector was flushed with pure CO$_{2}$ having a flow rate of 3.0 L/h for 4 hours prior to powering the detector. Using the power supply (CAEN-N1470), the voltage across the detector was increased in the steps of 200 V initially upto 3000 V, and after this the voltage was increased in the steps of 100 V and the corresponding current was measured. Simultaneous measurement of signal counts was done for every set of voltage. The rate calculated from these counts is called spurious signal rate which is defined as the rate of signals which are not originated from the ionization of the gas. The rate of the spurious signal is measured from the bottom of the last GEM foil as a function of the current running through the resistive divider of the HV distribution circuit. 

\begin{figure}[!ht]
    \begin{subfigure}[b]{0.46\textwidth}
        \includegraphics[scale=0.20]{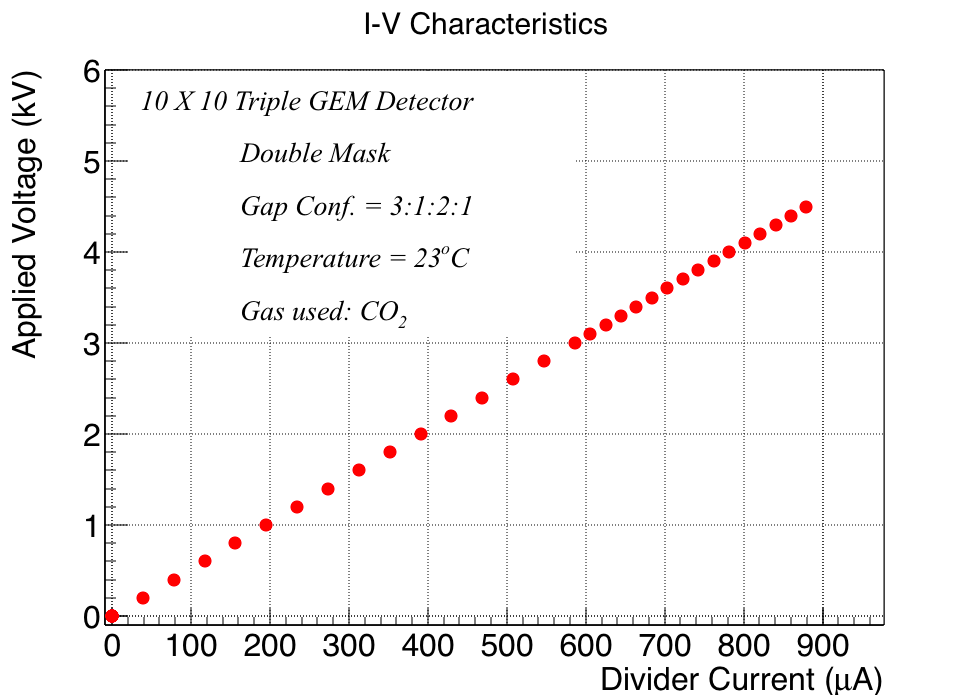}\qquad
        \caption{ }
    \end{subfigure}
    \hspace{0.5cm}
    \begin{subfigure}[b]{0.42\textwidth}
        \includegraphics[scale=0.20]{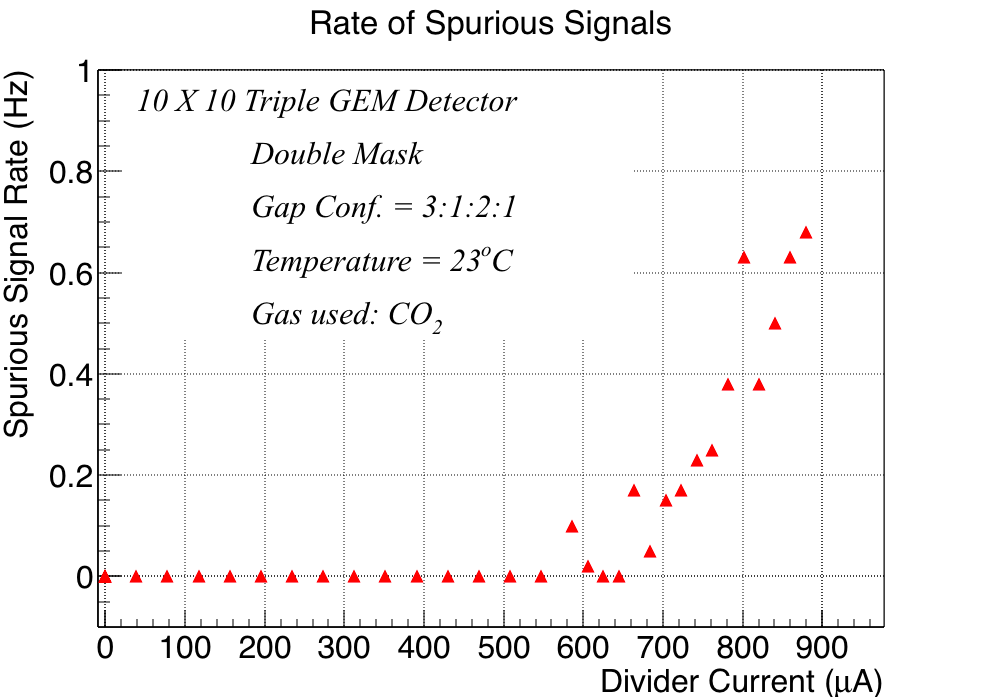}
        \caption{ }
    \end{subfigure}
   \caption{(a) Shows the I-V characteristics of the detector with an equivalent resistance of 5.115 $M\Omega$, and (b) shows the variation of spurious signal rate obtained from GEM3 bottom as a function of divider current.} 
\label{fig:HVTest}
\end{figure}

During the operation the detector shows an ohmic behaviour which can be clearly seen in fig.~\ref{fig:HVTest}(a). Thus, plotting I-V characteristics shows straight line behaviour and its slope gives us the total effective resistance of the circuit, which comes out to be 5.115 M$\Omega$. As shown in fig.~\ref{fig:HVTest}(b), the rate of the spurious signal shows random behaviour as we increase the voltage/current across the detector and the maximum spurious signal rate observed is 0.7 $Hz$ at a divider current of $\sim$900 $\mu A$.

\subsection{Gain measurement}

\begin{figure}[!ht]
\centering
\includegraphics[scale=0.27]{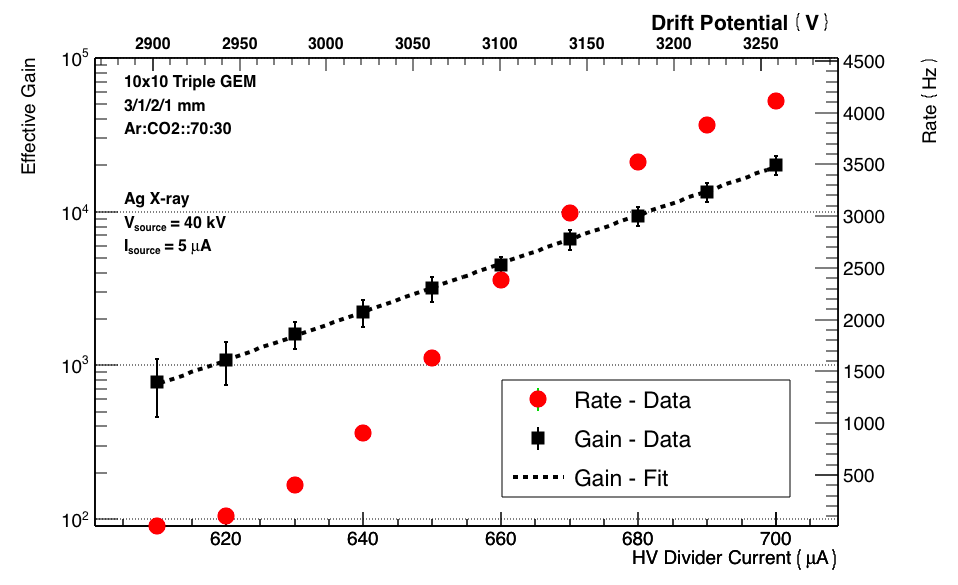}
\caption{The variation of rate and gain of triple GEM detector as a function of divider current and drift voltage. The measurements were taken at an ambient temperature of 21.9$^{o}C$ and pressure of 975.3 $hPa$.}
\label{fig:GainTest}
\end{figure}

\begin{linenomath*}
The effective gain is the central parameter of the GEM detector which explains the geometrical as well as the electrical properties together with the gas compositions. Also, in order to localize and detect particle efficiently, a stable and homogeneous gain over a long operation period is necessary. As explained in the Section 2, under the influence of the high electric field the electrons are guided into the GEM holes which leads to the multiplication of the charge. But, due to the fact that not all the amplified charge reach the anode, on the basis of which one can differentiate the ``\textit{real}'' gain with the ``\textit{effective}'' gain \cite{GainMes}. Effective gain is defined as the ratio of the charge arriving at the anode and the charge created from the primary ionisation in the drift field. It is expressed as:
\begin{equation}
G = \frac{\mid I_{with~source} - I_{without~source} \mid } {n_{primary} ~e ~R}
\end{equation}
where,  $I_{with~source}$ is the current measured from readout in the presence of X-ray source, $I_{without~source}$ is the current measured from readout in the absence of any source, $n_{primary}$ are the number of primary electrons created in the drift region which is $\sim$ 346 while operating X-rays at a voltage of 40 $kV$ and current of 5 $\mu A$, $e$ is the electronic charge and $R$ is the rate measured at the maximum divider current.
\end{linenomath*}

For the standard CMS GE1/1 GEM detector gap configuration, a gain of more than $10^4$ is measured at detector current of 700 $\mu A$. So, in order to start the gain measurement we also ramped up the detector to 700 $\mu A$. Then, the rate and the corresponding current are recorded both with and without X-ray source. After this, we decrease the current across the detector in the step of 10 $\mu A$ and again measure the readout current and the rate for both the settings of X-rays. This procedure is iterated until the rate with and without source becomes equal i.e. no ionization takes place in the chamber due to X-rays. As we observe in fig.~\ref{fig:GainTest}, the effective gain of the GEM detector exponentially increases with increase in the detector voltage. The maximum measured rate is $\sim$ 4 $kHz$; with corresponding measured gain of $\sim 2 \times 10^{4}$ at divider current of 700 $\mu A$. 

\subsection{Energy resolution}

\begin{linenomath*}
Energy resolution is ability of detector to precisely determine the energy of the incoming radiations. Energy resolution ($E_{res}$) of GEM detector for incoming X-ray source is defined as the percentage of Full Width Half Maximum (FWHM) to the height of the resulting Gaussian peak, and given as:
\begin{equation}
E_{res} = \frac{FWHM}{Peak ~Energy} = \frac{2.35 \times \sigma}{Peak~ Energy}
\end{equation}
and calculated using Gaussian fit parameters of the main peak. 
\end{linenomath*}

\begin{figure}[!ht]
\centering
\includegraphics[scale=0.2]{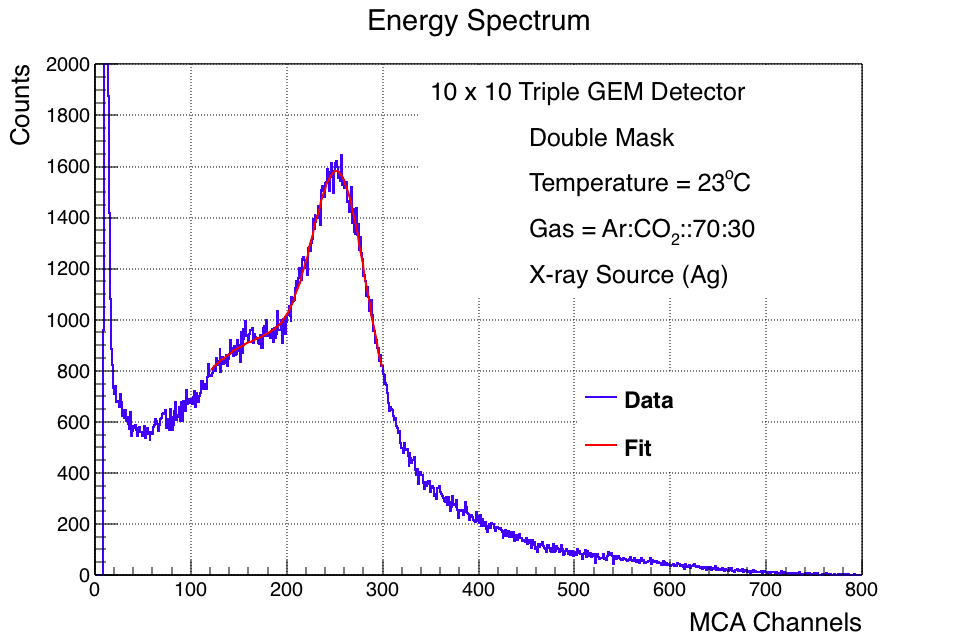}
\caption{Energy Spectrum obtained with X-rays (Ag-target) operated at 40 $kV$ and 5 $\mu A$ at divider current of 700 $\mu A$. These observations were taken at an ambient temperature of 23$^o C$, pressure of 970 $hPa$ and relative humidity of 32 \%.}
\label{fig:Spectrum}
\end{figure}

Figure~\ref{fig:Spectrum} shows a typical energy spectrum acquired with the silver target X-ray source. The energy resolution as calculated from the peak is $\sim$25$\%$ at the divider current of 700 $\mu A$ across the detector. 

\begin{figure}[!ht]
    \begin{subfigure}[b]{0.46\textwidth}
        \includegraphics[scale=0.325]{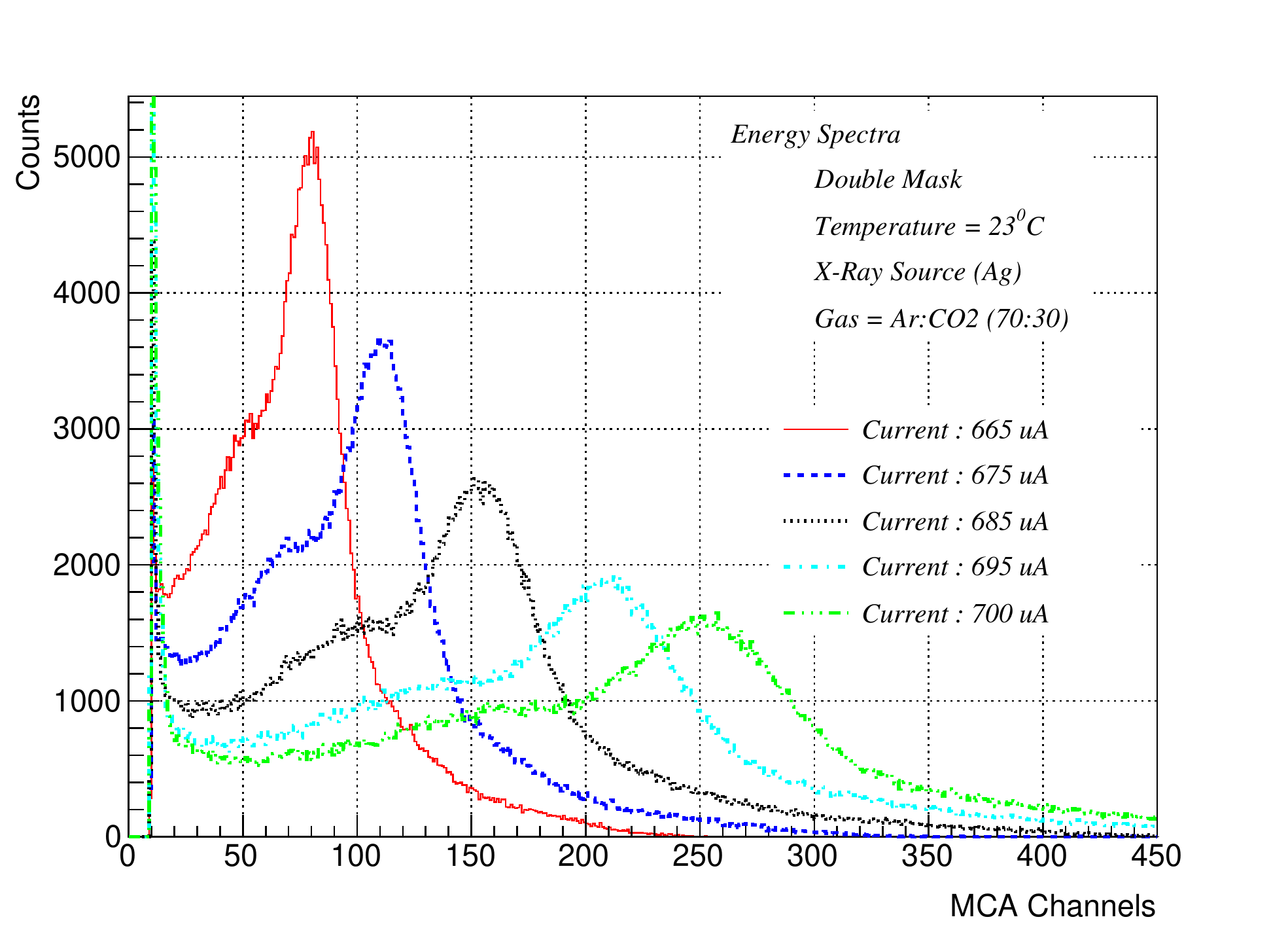}\qquad
        \caption{ }
    \end{subfigure}
    \begin{subfigure}[b]{0.46\textwidth}
        \includegraphics[scale=0.20]{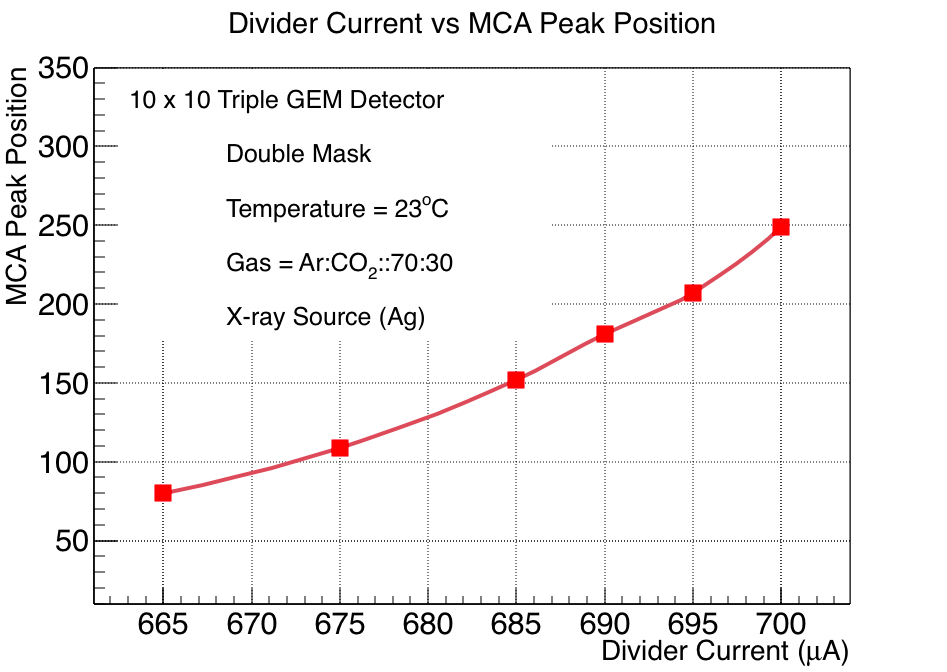}
        \caption{ }
    \end{subfigure}
   \caption{(a) Pulse height spectra for various GEM voltages, and (b) variation of MCA peak position as a function of divider current across the detector. These observations were taken at an ambient temperature of 23$^o C$, pressure of 970 $hPa$ and relative humidity of 32 \%.} 
\label{fig:Superimpose}
\end{figure}

As in the case of gain measurement of the detector, we again decrease the value of divider current in the steps of 5 or 10 $\mu A$ and plotted the corresponding energy spectra (as shown in fig. ~\ref{fig:Superimpose}(a)). As we decrease the current across the detector, the peak position of MCA shifts towards the left which explains that the avalanche electrons decreases with the decrease in the voltage across the GEM detector \cite{ResVar}. Figure ~\ref{fig:Superimpose}(b) shows the peak position of all the spectra with respect to the divider current which shows an exponential behaviour as expected.

\subsection{Gain uniformity}
There are several factors responsible for non-uniformity of gain over the whole surface of the GEM detectors. This includes a variation of gas gaps due to inaccurate stretching, quality of GEM foils, non-uniformity or defects in holes over the foil area, etc. Thus, the gain uniformity test is quite significant for checking the detector response over the entire surface area. For this, we have divided the active region of the detector i.e. $100~cm^{2}$ into 5 $\times$ 5 sectors of equal area. The X-ray source with a collimator was then placed at the centre of each sector and at a particular value of divider current the gain at each sector was measured, as explained in Section 4.2. 

\begin{figure}[!ht]
    \begin{subfigure}[b]{0.46\textwidth}
        \includegraphics[scale=0.225]{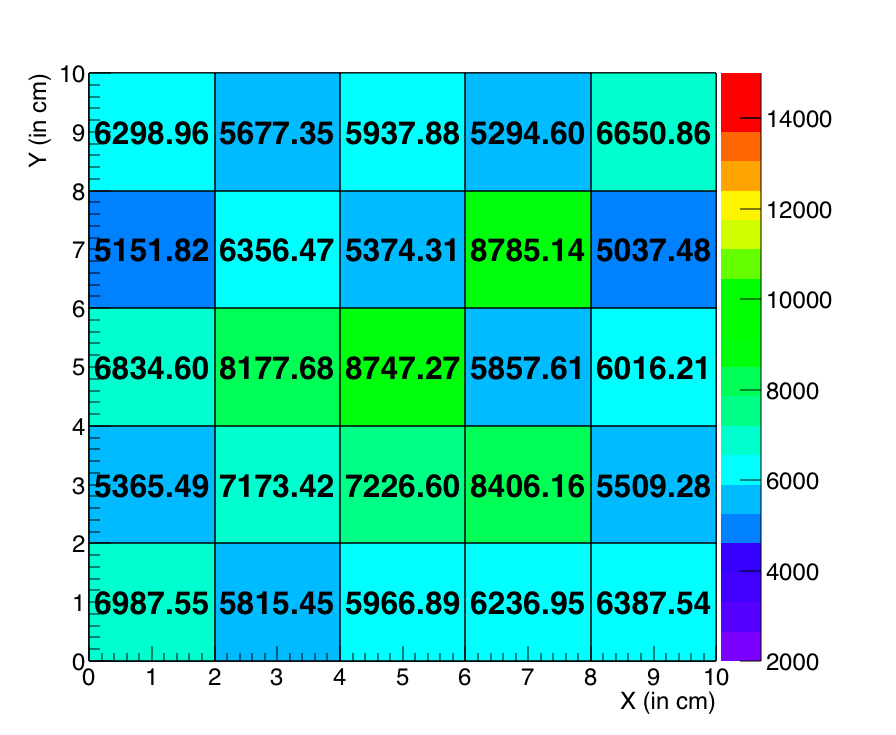}\qquad
        \caption{ }
    \end{subfigure}
    \hspace{0.5cm}
    \begin{subfigure}[b]{0.46\textwidth}
        \includegraphics[scale=0.225]{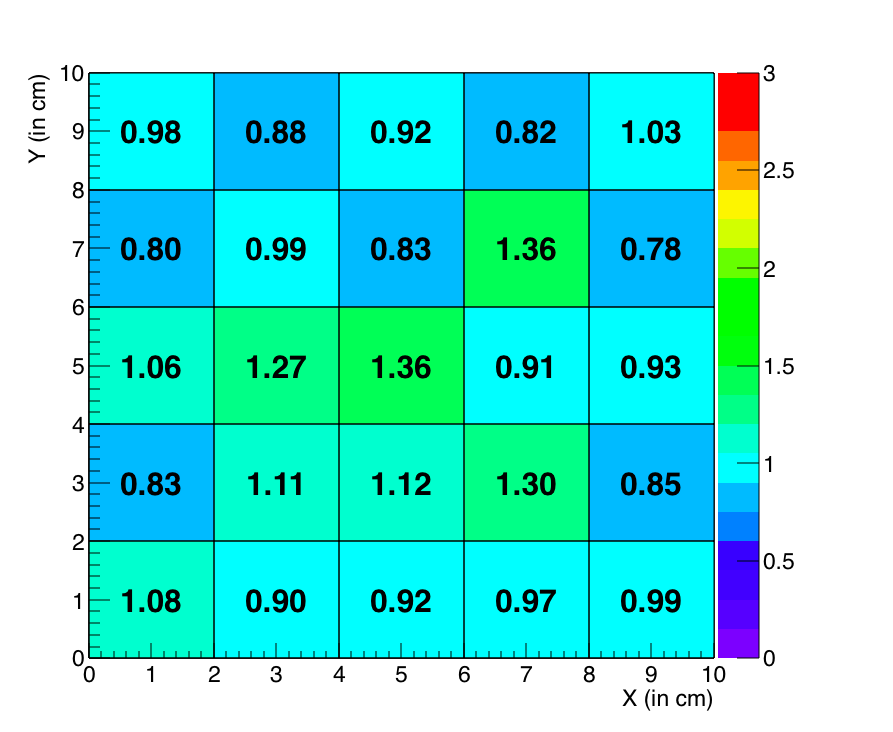}
        \caption{ }
    \end{subfigure}
   \caption{(a) Uniformity of gain for 5 $\times$ 5 equal sectors of $10~cm ~\times ~10~cm$ triple GEM detector, and (b) normalized gain uniformity with respect to the average value of gain. These observations were taken at an ambient temperature of 25$^o C$, pressure of 971 $hPa$ and relative humidity of 28\%.} 
\label{fig:GainUnif}
\end{figure}

Figure~\ref{fig:GainUnif}(a) shows the effective gain value for each sector of the GEM detector at divider current of 650 $\mu A$. As we can see, the effective gain is higher in the central region of the detector and lower at the corners. This is due to the fact that the probability is high for electrons to escape the detector when they are getting multiplied at the corners as compared to the central part of the detector \cite{MerlinThesis}. Figure~\ref{fig:GainUnif}(b) shows the normalized plot of the gain uniformity in which gain value in each sector is divided by the average gain over the entire surface of the detector. This figure also shows the same behaviour i.e. higher values at centre and lower at the edges.

\subsection{Correlation between foil defects and uniformity}
The operational characteristics of a GEM foil can vary with local variation in the size or the shape of the holes. Therefore, the distribution and size of the holes over a GEM foil should be uniform to achieve uniform functionality of the detector over the active surface. Process of manufacturing GEM foils can introduce occasional defects which may affect the performance of a GEM detector. During optical scanning of these commercially manufactured GEM foils, different types of defects were observed like over-etched, under-etched, unetched, burnt, merged holes etc., which were mainly due to the etching process \cite{Foils}. These are broadly divided into two types: Insulator defects and Copper defects. Since the gain of the detector also depends on the geometry of the foils and the defects present on it \cite{Correlation}, we have prepared a defect map for each foil showing the position of various defects present on it. This is performed by dividing the foil into 5 $\times$ 5 sectors of equal area so that we can see one to one correspondence of the defects with the results of uniformity. 

\begin{figure}[!ht]
\centering
    \begin{subfigure}[b]{0.46\textwidth}
        \includegraphics[scale=0.27]{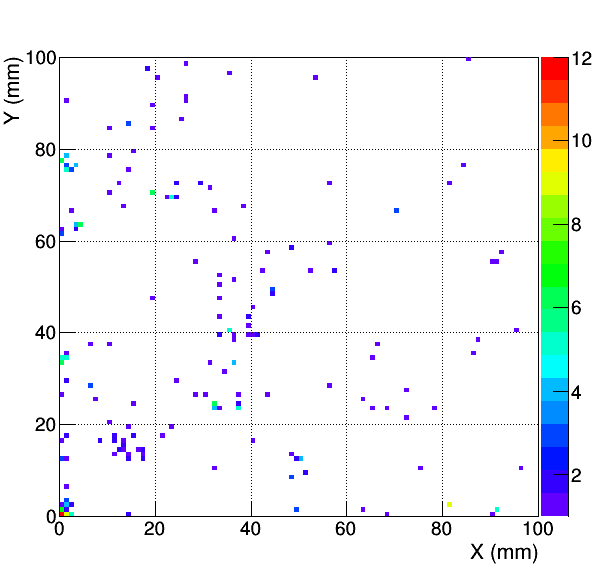}\qquad
        \caption{ }
    \end{subfigure}
    \begin{subfigure}[b]{0.46\textwidth}
        \includegraphics[scale=0.27]{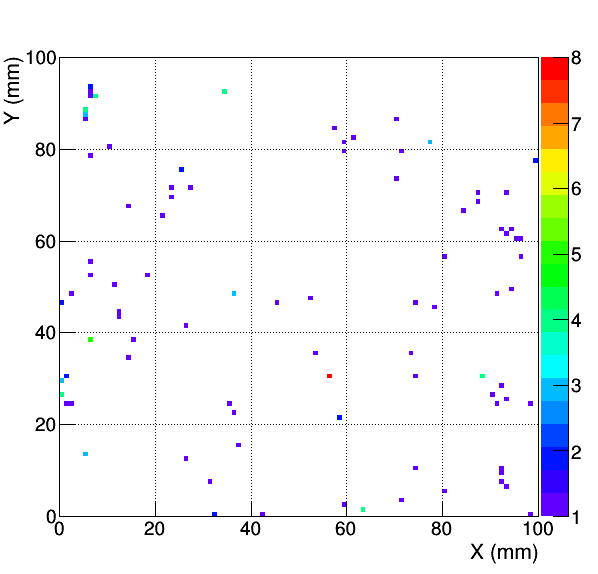}\qquad
        \caption{ }
    \end{subfigure}
    \begin{subfigure}[b]{0.46\textwidth}
        \includegraphics[scale=0.27]{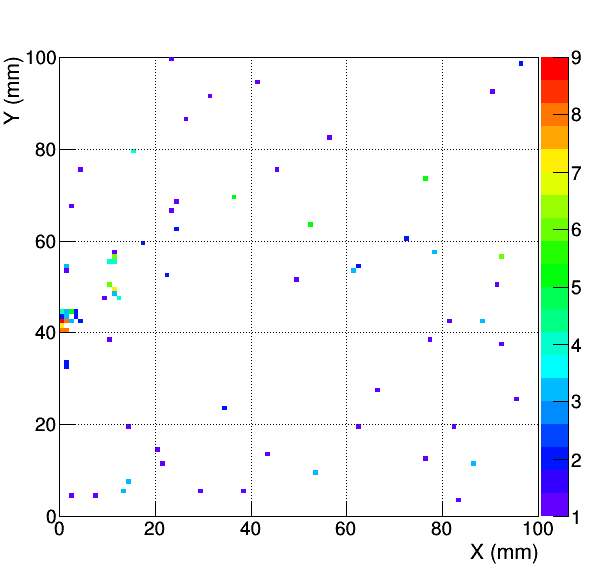}\qquad
        \caption{ }
    \end{subfigure}   
\caption{Showing position of all types of defects present in (a) first GEM foil, (b) second GEM foil, and (c) third GEM foil, used in the assembly of the GEM detector.}
\label{fig:Defects}
\end{figure}

\begin{figure}
\centering
\includegraphics[scale=0.27]{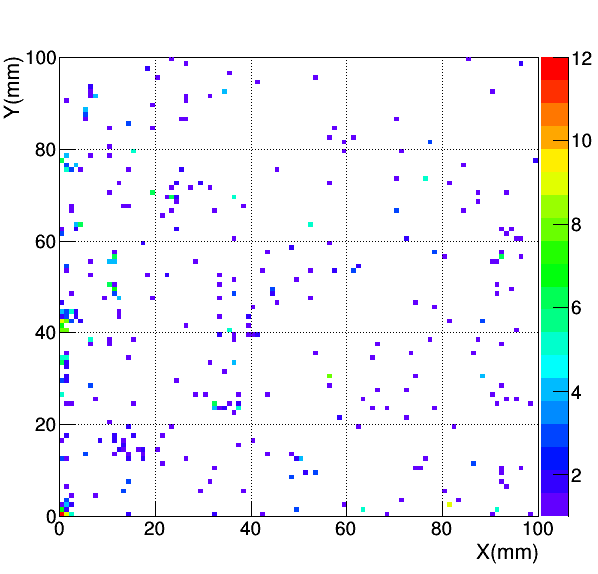}
\caption{Total defects after stacking the three foils inside the detector.}
\label{fig:DDefects}
\end{figure}

Since the defects present in the foils are of the order of $\mu m$ so to make them visible in the figure each defective hole is enlarged 20 times ($\sim$1 $mm$ in the figure). Also, each figure includes the defects from both the sides of the GEM foil as well as both types of defects (i.e Insulator and Copper defects). Figure~\ref{fig:Defects} shows the position of all defects of three foils. We have also prepared a defect map, as shown in fig.~\ref{fig:DDefects}, summarizing total defects of these foils according to the stack orientation of the foils inside the detector. As seen from the figure, the defects due to all three foils were distributed over the entire area of the detector. In order to estimate the effect of defects, we need to compare fig.~\ref{fig:DDefects} with the uniformity shown in fig.~\ref{fig:GainUnif}. Even though the defects are distributed over the entire foil(s) but if we look closely then some sectors of the foil(s) are having some more defects than others. For e.g., the total defects in ($x,~y$) (0-20$mm$, 0-20$mm$) sector are comparatively more but the same sector in fig.~\ref{fig:GainUnif} does not show any reduction in gain. The gain in other sectors also does not show any direct correlation with defect map. This may be due to the fact that the actual number of defective holes are quite small compared to good holes (i.e. 785 defective holes in 600,000 holes which corresponds to 0.13\% of defects) and hence it does not affect the performance of the foil(s). 

\section{Result}
Various studies have been performed on triple GEM detector assembled using first India-made $10~cm ~\times ~10~cm$ GEM foils. No sparks and voltage trips were observed in the detector under the high voltage upto 4.9 $kV$ in pure CO$_{2}$ medium and the effective resistance obtained from I-V characteristics is 5.115 $M\Omega$ which accounts to less than 5 \% of error to actual value. The spurious signal rate was not much significant and found to be less than 1 $Hz$ at the current across the detector being 900 $\mu A$. Using X-rays, a gas gain of $\sim$ 20$k$ have been obtained at a divider current of 700 $\mu A$ with the gas mixture of Ar:CO$_{2}$ in 70:30 proportions. Energy spectrum is also obtained under the same conditions and energy resolution was calculated to be 25\% for divider current varying between 660-700 $\mu A$. By dividing the active area of the detector into 5 $\times$ 5 equal sectors, uniformity is plotted and compared with the defect maps for individual foils as well as the stack orientation. By comparing the results of optical scanning with local gain measurement across the detector, a distinct correlation was found. 

\section{Conclusions}
Micropack successfully built the small-area GEM foils, the foils are tested for their detector performance and compared with the one produced at CERN, which shows similar response.
These measurements are going to be very beneficial for future development since the commercially manufactured large area GEM foils are needed in  future upgrades of CMS Detector i.e. GE2/1 and ME0. These foils can also be used by other experiments and technology can be exploited in inter disciplinary applications such as medical imaging, etc.

\section{Acknowledgements}
We would like to acknowledge the funding agency, Department of Science and Technology (DST), New Delhi (grant nos. SR/MF/PS-02/2014-DUA and SR/MF/PS-02/2013-DUN) for providing financial support.

\end{document}